\documentclass[intlimits,twoside,a4paper]{article}

\usepackage{amsmath,amssymb}
\usepackage{graphicx}

\usepackage[T2A]{fontenc}
\usepackage[cp1251]{inputenc}

\usepackage[eqsecnum]{cmpj3}

\issue{2017}{20}{4}{43701}
\doinumber{10.5488/CMP.20.43701}
\title{Integer quantum Hall effect and topological phase transitions in silicene}
\author[Y.L. Liu \textsl{et al.}]{Y.L. Liu\refaddr{label1}, G.X. Luo\refaddr{label2}, N. Xu\refaddr{label2}\thanks{xuning79530@126.com}\,,  H.Y. Tian\refaddr{label4}, C.D. Ren\refaddr{label3}\thanks{renchongdan@hotmail.com}}
\addresses{
\addr{label1} Faculty of Electric Information Engineering, Huaiyin Institute of Technology, Huaian 223001, China
\addr{label2} Department of Physics, Yancheng Institute of Technology, Jiangsu 224051, China
\addr{label4} School of Physics and Electronic Engineering, Linyi University, Linyi 
 276005, China
\addr{label3} Department of Physics, Zunyi Normal College, Guizhou 563002, China
}

\date{Received January 15, 2017, in final form April 23, 2017}
\begin{document}

\maketitle

\begin{abstract}
We numerically investigate the effects of disorder on the quantum Hall effect (QHE) and the quantum phase
transitions in silicene based on a lattice model.
It is shown that for a clean sample, silicene exhibits an unconventional QHE near the band center, with plateaus developing at
$\nu=0,\pm2,\pm6,\ldots,$ and a conventional QHE near the band edges.
In the presence of disorder, the Hall plateaus can be destroyed
through the float-up of extended levels toward the band center, in which higher plateaus disappear first.
However, the center $\nu=0$ Hall plateau is more sensitive to disorder and disappears at a relatively weak disorder strength.
Moreover, the combination of an electric field and the intrinsic spin-orbit interaction (SOI) can lead to quantum phase
transitions from a topological insulator to a band insulator at the charge neutrality point (CNP), accompanied by
additional quantum Hall conductivity plateaus.
\pacs 73.43.-f, 73.43.Nq, 72.80.Ey
\end{abstract}

\vspace{-5mm}
\section{Introduction}
\begin{figure}[!t]
\begin{center}
\includegraphics[scale=0.5]{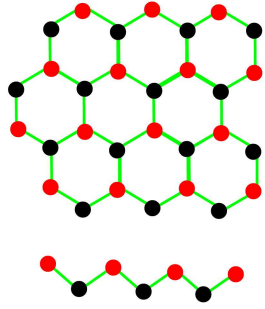}
\end{center}
\caption{\label{fig1}(Color online) Illustration of buckled silicene.
The upper and lower images correspond to top and side views, respectively, of the lattice geometry.
Due to the large ionic size of
silicon atoms, the silicene lattice is buckled, and the two sublattices
sit in vertically separated, parallel planes.}
\end{figure}

Silicene is a monolayer of silicon atoms
bonded together on a two-dimensional (2D) honeycomb
lattice. Both silicene sheets and ribbons
have been experimentally synthesized through synthesis on metal surfaces \cite{Majzik.2013,Padova.2010,Vogt.2012,Chiappe.2014}. Silicene
shares almost every remarkable property of graphene; for instance,
it exhibits Dirac-like electron dispersion at the corners of the
Brillouin zone.
Unlike graphene, it has a buckled structure due to
the large ionic radius of silicon atoms \cite{Drummond.2012,Liu.2011,Liu.2011b}, which causes different
sublattices to sit in different vertical planes with a separation
of $d\thickapprox0.46$~{\AA} \cite{Drummond.2012,An.2012}, as shown in figure~\ref{fig1}. When an electric field $E_z$ is
applied perpendicular to the silicene plane, an on-site potential
difference ($\Delta_z=E_zd$) will be created between  different
sublattices \cite{Drummond.2012,Ni.2012,Ezawa.2012}.
Moreover, a buckled structure leads to a relatively large spin-orbit gap of
$\Delta_{\text{so}}\approx1.55{-}7.9$~meV \cite{Drummond.2012,Liu.2011,Liu.2011b,An.2012}, as obtained through first-principles calculations
and tight-binding calculations \cite{Liu.2011b}, which provides a
mass to the Dirac fermions.
When the strength of $\Delta_z$ becomes greater than
$\Delta_{\text{so}}$, silicene will undergo a transition from a topological insulator (TI) to a band
insulator (BI) \cite{Drummond.2012,Ezawa.2012,Tabert.2013}.

Under a perpendicular magnetic field $B$, graphene exhibits a $\sqrt{B}$-dependent Landau level (LL) spectrum and gives rise to an unusual
integer quantum Hall effect (QHE).
Similar to graphene, the $n\neq 0$ LLs of silicene scale as $\sqrt{B}$,
whereas the $n=0$ LL is not pinned at zero energy as in graphene \cite{Tabert.2013,Ullah.2014}.
Therefore, silicene is expected to exhibit an exotic QHE, and the $n=0$ LL
should possess peculiar properties under the combination of a perpendicular electric field and the spin-orbit interaction (SOI).
Recently, the QHE and quantum phase
transitions of silicene have been investigated by many authors \cite{Ezawa.2012b,Tahir.2013,Shakouri.2014a,Shakouri.2014b,Shakouri.2014};
however, studies of the effect of disorder on the QHE in silicene,
which is an essential ingredient for understanding the underlying physics of the
QHE phenomenon compared with that in graphene \cite{Sheng.2006}, are still lacking.
Therefore, it is highly desirable to numerically investigate the
effect of disorder on the QHE in silicene while considering the full band structure
to reveal the underlying physics of the QHE phenomenon.

In this work, we conduct a numerical study of the QHE
in silicene in the presence of disorder and an electric field based on a
tight-binding model.
It is shown that unconventional QHE plateaus at $\nu=0,\pm2,\pm6,\ldots$ are produced near the band
center and conventional QHE plateaus appear near the
band edges.
We further map out the phase diagram for the
QHE and demonstrate that the Hall plateaus disappear
at strong disorder through the float-up of extended levels toward the band center.
However, the $\nu=0$ plateau is not as stable as other plateaus with the same plateau width near the band center
and will disappear at a weak disorder.
Under a perpendicular electric field, the spin and valley degeneracies
are lifted, and the system changes from the TI phase to BI phase
when the strength of the electric field exceeds the strength of the SOI.

\section{Model and formalism}
We consider a buckled silicene structure on a honeycomb lattice,
where the A and B sublattices sit in different vertical planes,
as shown in figure~\ref{fig1}. The entire system contains $L_y$ zigzag chains with
$L_x$ atoms in each chain, and the size of the sample could be expressed as
$N=L_x\times L_y$. Under a perpendicular electric field $E_z$,
a staggered potential $\Delta_z=E_zd$ is generated between  different
sublattices. In the presence of both a magnetic
field and an electric field applied perpendicular
to the silicene plane,
the lattice model can be written in the tight-binding form as follows \cite{Liu.2011b}:
\begin{eqnarray}
\label{eq1}
H=\sum_{i\alpha}(\epsilon_{i\alpha}+w_{i\alpha})
a^{\dagger}_{i\alpha}a_{i\alpha}-t\sum_{\langle i,j\rangle\alpha}\re^{\ri\phi_{ij}}a^{\dagger}_{i\alpha}a_{j\alpha} 
+\ri\frac{\lambda_{\text{so}}}{3\sqrt{3}}\sum_{\langle\langle i,j\rangle\rangle\alpha\beta}\nu_{ij}a_{i\alpha}^
{\dagger}\sigma_{\alpha\beta}^{z}a_{j\beta}+\Delta_z
\sum_{ i\alpha}a_{i\alpha}^{\dagger}\tau_{z}a_{i\alpha}\,,
\end{eqnarray}
where $a_{i\alpha}^{\dagger}$ and $a_{i\alpha}$ are the creation
and annihilation operators
at the $i$-th discrete site with spin polarization~$\alpha$ and
the $\langle i,j\rangle$ ($\langle\langle i,j\rangle\rangle$)
run over all nearest-neighbor (next-nearest-neighbor) hopping sites.
The first term represents
the on-site energy and random disorder, where the on-site
disorder energy $w_i$
is uniformly distributed in the range $w_i\in[-W/2, W/2]t$ in terms of the
 disorder strength $W$.
The second and third terms represent the nearest and
next-nearest couplings with the effective SOI $\lambda_{\text{so}}$, where $\sigma=(\sigma_x,\sigma_y,\sigma_z)$ is
the Pauli spin matrices and $\nu_{ij}=+1$ and $\nu_{ij}=-1$
correspond to anticlockwise and clockwise
next-nearest-neighboring hopping, respectively, with respect to the positive
$z$ axis.
The final term represents the lattice potential
resulting from different sublattices in the perpendicular electric
field $E_z$, where $\tau_z$ is the Pauli matrix of the sublattice.
In the presence of a
magnetic field $B$, a Peierls phase factor $\phi_{ij}$ is
added to the hopping interactions,
where $\phi_{ij}=\int_{i}^{j}\vec{A}\cdot \rd\vec{l}/\phi_{0}$,
with the vector potential
$\vec{A}=(-By,0,0)$ and $\phi_0=\hbar/e$. The magnetic flux
per hexagon is $\phi=2\piup/M$,
where $M$ is an integer. The total flux through the
sample, $N\phi/4\piup$, is taken to be an integer. When $M$
is equal to $L_x$
or $L_y$, the magnetic
periodic boundary conditions are reduced to ordinary
periodic boundary conditions.
Here, we disregard the Rashba SOI because it is weak in comparison with
the intrinsic SOI \cite{Liu.2011b}.

In the presence of an
electric field, the spin degeneracy is broken; therefore,
it is convenient to investigate the Hall conductivity
separately for each spin.
We use the symbol $s$ $(s=\uparrow,\downarrow)$ to define
the spin up or spin down.
The eigenstates $|\alpha^s\rangle$ and eigenenergies
$\epsilon_{\alpha}^{s}$ for each spin are
obtained through exact diagonalization of the Hamiltonian in
equation~(\ref{eq1}), and the Hall conductivity for different
spins $\sigma_{xy}^{s}$ is calculated
using the Kubo formula \cite{Sheng.2006}:
\begin{eqnarray}
\sigma_{xy}^s=\frac{\ri e^2\hbar}{S}\sum_{\alpha,\beta}\frac{\langle\alpha^s| V_x^s|\beta^s\rangle\langle\beta^s| V_y^s|\alpha^s\rangle-\text{h.c.}}
{(\epsilon_{\alpha}^{s}-\epsilon_{\beta}^{s})^2}\,,
\end{eqnarray}
where $S$ is the area of the sample and $V_x^s$
and $V_y^s$ are the velocity operators of spin $s$. The total
Hall conductivity of the system is $\sigma_{xy}=\sigma_{xy}^{\uparrow}+\sigma_{xy}^{\downarrow}$.
 It is hoped that our results will suggest new potential
directions for the experimental realization of the QHE and topological insulators.

\section{Results and discussion}
\begin{figure}[!b]
\vspace{-7mm}
\begin{center}
\includegraphics[scale=0.4]{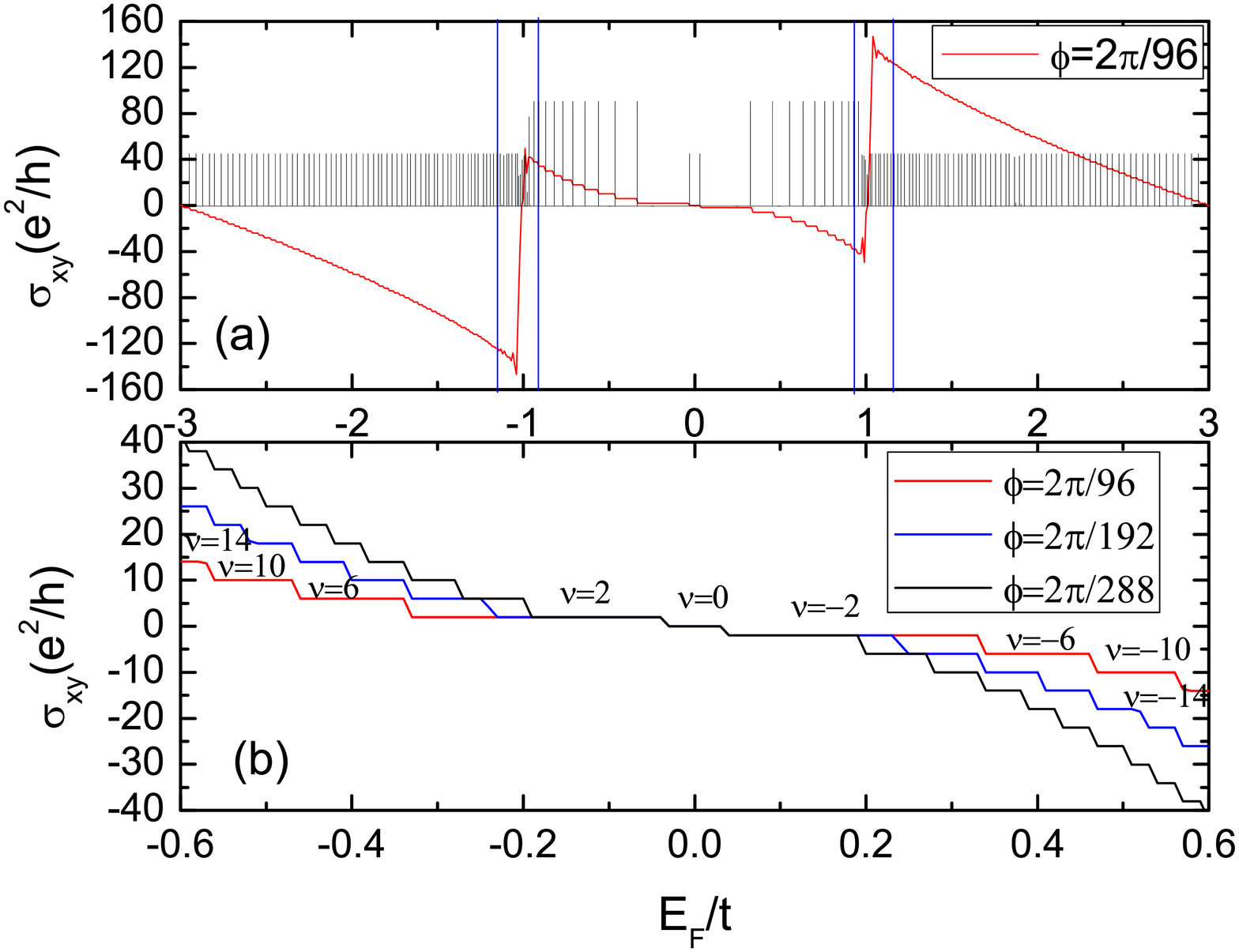}
\end{center}
\vspace{-5mm}
\caption{\label{fig2} (Color online) (a) The Hall conductivity $\sigma_{xy}$ and electron density of states (in units of 1)
throughout the entire energy band with a magnetic flux of $\phi=2\piup/96$.
(b) The Hall conductivity near the band center for different magnetic fluxes.
The disorder strength is $W=0$, and the electric field strength is $\Delta_z=0$;
the system size is $N=288\times32$.}
\end{figure}

In what follows, we will discuss the Hall conductivity $\sigma_{xy}$
and the quantum phase transitions in silicene. In our calculations,
the hopping integrals and the SOI strength
are taken to be $t=1.6$~eV and $\lambda_{\text{so}}=0.04t$, respectively; the temperature
is set to $T=0$~K, and the disorder is averaged
over 800 sample configurations.

We first focus on the Hall conductivity and the electron density
of states as functions of the electron Fermi energy $E_{\text f}$ for
a clean sample ($W=0$) under zero electric field; the
results are shown in figure~\ref{fig2}. The system size is taken to be
$N=288\times32$ and the magnetic flux is taken to be $\phi=2\piup/96$ (corresponding to approximately 822~T \cite{Cresti.2008}) to illustrate
the overall picture of the QHE throughout the entire energy band.

It is shown in figure~\ref{fig2}~(a) that discrete LLs appear in the
system and constitute a nonzero density of states.
The LLs near the band center are four-fold degenerate due to the two spin components and two Dirac points,
except for the two center $n=0$~LLs adjacent to the gap, which show a spin splitting: the upper one is the $n=0$
spin-up LL, and the lower one is the $n=0$ spin-down LL. The LL above (below) the
spin-up (spin-down) $n=0$ LL is the $n=1$ ($n=-1$)~LL, and so on.
Therefore, at the band center, the quantum Hall conductivity $\sigma_{xy}=\nu e^2/h$ is quantized according to
the unconventional quantization rule $\nu=g_s(k-1/2)e^2/h$ $(k=1,2,\ldots)$
beyond the gap, where $g_s=4$ is the LL degeneracy.
However, a zero Hall plateau appears in the energy gap ($-\lambda_{\text{so}}\leqslant E_{\text f}\leqslant\lambda_{\text{so}}$)
because the two $n=0$~LLs reside at $E_{\text f}=\pm\lambda_{\text{so}}$
and each contributes a Hall conductivity of $\pm e^2/h$, respectively, to $\sigma_{xy}$.
A similar zero quantum Hall plateau induced by the quantum spin Hall (QSH) gap can also appear in graphene under a strong magnetic field;
however, it is induced by Zeeman splitting rather than by the SOI \cite{Young.2014}.
In figure~\ref{fig2}~(b), the quantization rule for the Hall conductance in this
unconventional region is shown for three different $\phi$.
It is shown that more quantum plateaus appear with a decreasing $\phi$, but
the $\nu=0$ plateau is not affected by a change in the magnetic field.

As each additional LL
is occupied, the total Hall conductivity  increases by $g_se^2/h$
as a result of the spin and valley degeneracy.
When the energy reaches $E_{\text f}=\pm t$, which are the van Hove singularities,
the Hall plateaus become indiscernible and the system changes to exhibit the
conventional QHE as it exceeds the critical region, as indicated by the
two blue lines. Similar to those in graphene \cite{Sheng.2006,Hatsugai.2006}, these crossover regions correspond to a transport
regime, in which the Hall resistance changes sign and the longitudinal
conductivity exhibits a metallic behavior.
At the band edges, the LLs are two-fold degenerate due to the two spin components,
and a conventional QHE emerges, with the Hall
conductivity quantized as $\sigma_{xy}=kg_se^2/h$ $(k=1,2,\ldots)$, where
$g_s=2$ because there is a spin degeneracy only.

\begin{figure}[!b]
\vspace{-2mm}
\begin{center}
\includegraphics[scale=0.5]{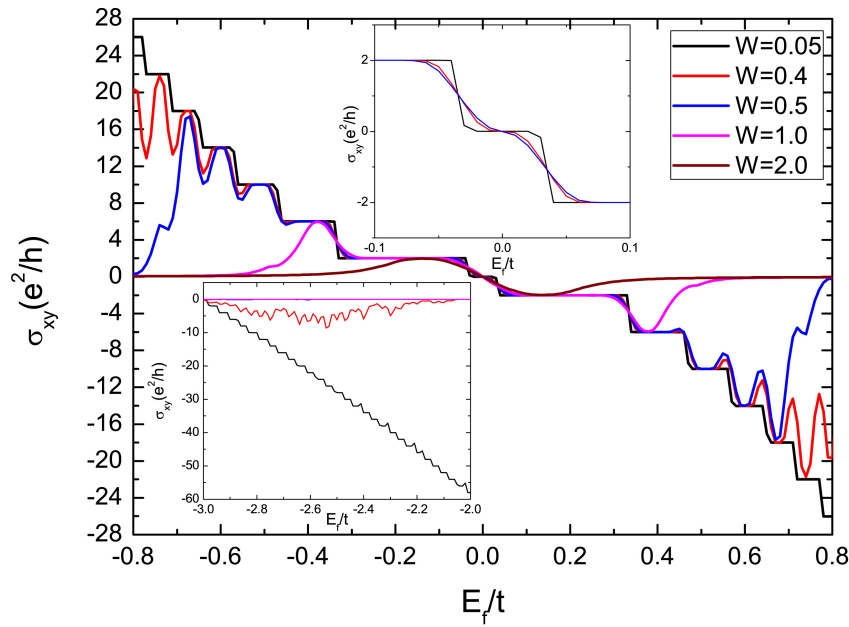}
\end{center}
\vspace{-5mm}
\caption{\label{fig3} (Color online) Calculated Hall conductivity $\sigma_{xy}$ as a
function of the electron Fermi energy for different disorder strengths.
The two insets show the Hall conductivity near the lower band edge and the central Hall plateau $\nu=0$.
The magnetic flux and system size are $\phi=2\piup/96$
and $N=288\times32$, respectively.}
\end{figure}

Now, we investigate the effect of random disorder on the QHE in silicene.
In figure~\ref{fig3}, the Hall conductance near the band
center is shown for different disorder strengths, with a magnetic flux of
$\phi=2\piup/96$ and a system size of $N=288\times32$.
It is shown that the higher plateaus (with larger~$\vert\nu\vert$) become less distinct and eventually disappear
with an increasing $W$.
At $W=0.05$, the plateaus at $\nu=0,\pm2,\pm6,\pm10,\pm14$ remain well
quantized, whereas all plateaus except the $\nu=\pm2$ plateau disappear at $W=2.0$.
By contrast, as shown in the lower inset of the panel,
for the QHE near the band edge, all plateaus are destroyed at a low disorder strength of $W=0.5$.
The center plateau ($\nu=0$) remains distinct at a small disorder strength of $W=0.05$;
however, it narrows at $W=0.4$ and
nearly disappears at a disorder strength of $W=0.5$
(the higher insert in figure~\ref{fig3}). Obviously, the $\nu=0$
plateau is not as robust as the one at $\nu=14$,
which has approximately the same initial width as the $\nu=0$ plateau
but still has a narrow, yet finite width at $W=0.5$.
This is because the magnetic field breaks the time-reversal symmetry
of the QSH states in the energy gap, and the $\nu=0$ states become unstable
under a weak disorder.

\begin{figure}[!b]
\vspace{-2mm}
\begin{center}
\includegraphics[scale=0.5]{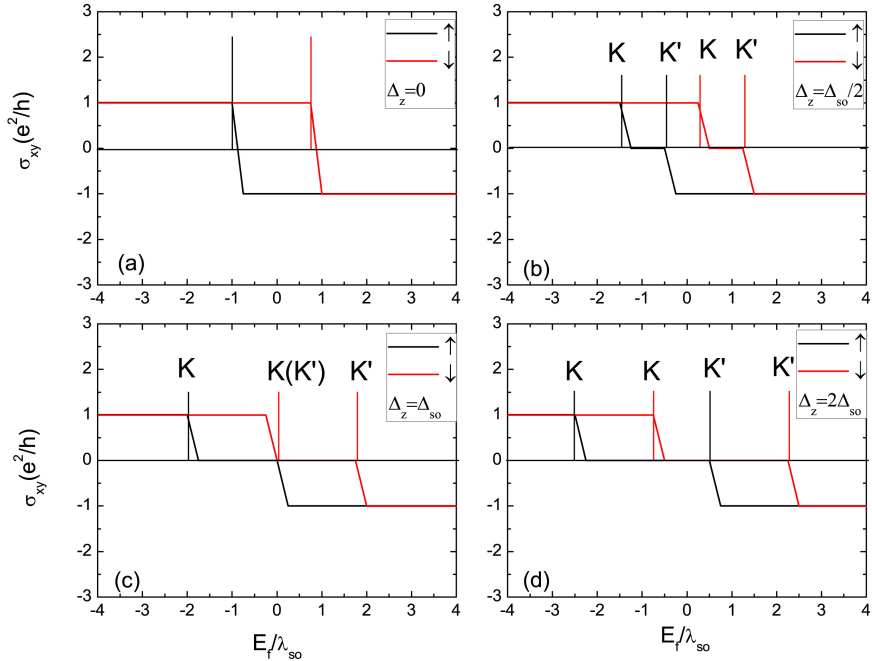}
\end{center}
\vspace{-5mm}
\caption{\label{fig4} (Color online) Spin-dependent Hall conductivity and electron density of states
of the $n=0$ states for different on-site potential differences $\Delta_z$. The black
and red lines correspond to the spin-down and spin-up states, respectively. The
notation K (K$'$) in the figure indicates the state from the K (K$'$) valley.
The magnetic flux and system size are $\phi=2\piup/96$
and $N=288\times32$, respectively.}
\end{figure}
Finally, we numerically investigate the quantum phase transition from a two-dimensional
TI to a trivial insulator under an external perpendicular electric field
with an on-site potential difference of $\Delta_z$; the results are shown in figure~\ref{fig4}. When $\Delta_z=0$,
the $n=0$ LLs are valley degenerate and spin split --- the spin-up LL is located at
$E_{\text f}=\lambda_{\text{so}}$, and the spin-down LL is located
at $E_{\text f}=-\lambda_{\text{so}}$.
They have the same density of states and make Hall conductivity contributions of
$\pm e^2/h$, leading to zero Hall conductivity
near the Dirac point.
In the presence of an
electric field, the valley degeneracy is resolved, which
results in four $n=0$ states; two are from the K valley, and two are from the K$'$
valley.
For a pair of spins from different valleys, there is a gap between them
that can be tuned by adjusting the electric field.
However, the gap between the two spins in each valley (K or K$'$) is fixed at the
SOI strength, $\lambda_{\text{so}}$.
As $\Delta_z$ increases, the valley split is enhanced; thus, the spin-up
state from the K valley descends and the spin-down state from the
K$'$ valley ascends. When $\Delta_z<\Delta_{\text{so}}$, the two states are
still above and below the Dirac point; therefore, the total system is still a
TI, and Hall plateaus appear at $\nu=0,\pm1,\pm2$ near the Dirac point,
as shown in figure~\ref{fig4}~(b).
At the critical point $\Delta_z=\Delta_{\text{so}}$,
the two states coincide at $E_{\text f}=0$ and the system transits into a
semi-metallic state, with Hall plateaus appearing at $\nu=0,\pm2$ near the Dirac point,
as shown in figure~\ref{fig4}~(c). When $\Delta_z>\Delta_{\text{so}}$,
band inversion occurs, with the spin-down
state from the K$'$ valley shifting from negative to positive
and the spin-up state from the
K valley shifting from positive to negative; the system
enters the BI regime, and Hall plateaus develop at $\nu=0,\pm1,\pm2$.
In figure~\ref{fig4}~(b) and (d), the $\nu=\pm1$ plateaus that are contributed by
one spin state from one valley arise only from the breaking of the valley degeneracy.
In fact, $\nu=\pm1$ Hall plateaus can also appear in graphene, in which case they originate
from the long-range Coulomb interaction \cite{Sheng.2007} or
the breaking of the spin and valley degeneracies \cite{Young.2014}.

\section{Conclusion}
In conclusion, we have numerically investigated the effects of disorder
on the QHE and the quantum phase transitions in silicene under a perpendicular
electric field.
It is shown that Hall plateaus develop at $\nu=0,\pm2,\pm6,\ldots$
near the band center, whereas conventional quantum Hall plateaus appear near the band edges.
The phase diagram indicates that the Hall plateaus gradually disappear
toward the band center with an increasing disorder strength,
which causes higher plateaus to disappear first.
However, the $\nu=0$ plateau is not robust and disappears at a relatively weak disorder strength.
In the presence of an electric field, the valley degeneracy of
the two $n=0$ LLs is broken, which results in a quantum phase transition at the CNP
together with the unconventional plateaus in the QHE.
The derived results also apply to isostructural
germanene, whose SOI strength is even higher ($\Delta_{\text{so}}=43$~meV with $d=0.66$~{\AA}).
Due to the strong SOI and electric field,
the QHE in silicene and germanene will be unaffected by temperature.
Therefore, the QHE and quantum phase transitions in silicene and
germanene can be experimentally observed and tuned at a finite temperature.
Electrical controllability in silicene and germanene is of great significance for the development of electrically tunable spintronic
and valleytronic devices.

\section*{Acknowledgements}
This work was supported by the NNSF of China (Nos. 11404278 and 11547189), the Natural Science Foundation of Jiangsu Province (NSFJS, No. BK20150423), the Science Foundation of Guizhou Science and Technology Department (No. QKHJZ[2015]2150), and the Science Foundation of Guizhou Provincial Education Department (No. QJHKYZ[2016]092).

\ukrainianpart
\title{ Цілочисловий квантовий Голів ефект та топологічні фазові переходи у сіліцені }
\author{Я.Л. Ліу\refaddr{label1}, Г.Х. Луо\refaddr{label2}, Н. Сюй\refaddr{label2},  Ґ.Я. Тіань\refaddr{label4}, Ц.Д. Рен\refaddr{label3}}
\addresses{
\addr{label1} Факультет електричної інформаційної техніки, Технологічний інститут Хуаинь, Хуайань 223001, Китай
\addr{label2} Відділ фізики, Технологічний інститут Яньчен, Цзянсу 224051, Китай
\addr{label4} Інститут фізики та електронної інженерії, Університет Ліньї, Ліньї 276005, Китай
\addr{label3} Відділ фізики, Нормальний коледж Цзуньї, Ґуйчжоу 563002, Китай
}

\makeukrtitle

\begin{abstract}
Проведено числове дослідження впливу безладу на квантовий Голів ефект (КГЕ) та квантові фазові переходи у сіліцені на основі ґраткової моделі.
Показано, що у випадку чистого зразка, сіліцен проявляє нетрадиційний КГЕ поблизу центру зони, де утворюються плато при 
$\nu=0,\pm2,\pm6,\ldots,$ і традиційний КГЕ поблизу країв зони. При наявності безладу,  плато Гола можуть бути зруйновані за рахунок
спливання розтягнутих рівнів в напрямку до центру зони, де першими зникають вищі плато. 
Однак, центр $\nu=0$ плато Гола 
є більш чутливим до безладу і зникає при відносно слабій силі безладу. Крім того, поєднання електричного поля та властивої спін-орбітальної взаємодії  може призвести до квантових фазових переходів від топологічного діелектрика до зонного діелектрика у точці нейтральності заряду, що супроводжується утворенням додаткових плато квантової провідності Гола.

\keywords квантовий Голів ефект, сіліцен, квантові фазові переходи
\end{abstract}

\end{document}